\documentclass{aa}
\usepackage{graphicx}
\newcommand{\be}{\begin{equation}}
\newcommand{\ee}{\end{equation}}
\begin{document}

\title{Propagation velocities of gas rings in collisional ring galaxies}

\author{E. I. Vorobyov\inst{1} \and D. Bizyaev\inst{2}}

\offprints{E. I. Vorobyov}

\institute{Institute of Physics, Stachki 194, Rostov-on-Don, Russia, and Isaac
Newton Institute of Chile, Rostov-on-Don Branch \\
\email{eduard\_vorobev@mail.ru}
\and
Sternberg Astronomical Institute, Universitetsky prospect 13, Moscow,
Russia; Isaac Newton Institute of Chile, Moscow Branch; Department of
Physics, UTEP, El Paso, Texas, USA \\
\email{dmbiz@sai.msu.ru}
}

\date{}

\abstract{The propagation velocity of the first gas ring in collisional ring
galaxies, i.e. the velocity at which the maximum in the radial gas
density profile propagates radially in the galactic disk, is usually inferred from
the radial expansion velocity of gas in the first ring.
Our numerical hydrodynamics modeling of ring galaxy formation however shows
that the maximum radial expansion velocity of gas in the first ring ($v_{\rm gas}$)
is invariably below the propagation velocity of the first gas ring itself ($v_{\rm
ring}$). Modeling of the Cartwheel galaxy indicates that the outer ring is
currently propagating at $v_{\rm ring} \approx 100$~km~s$^{-1}$, while the maximum
radial expansion velocity of gas in the outer ring is currently
$v_{\rm gas} \approx 65$~km~s$^{-1}$.
The latter value is in marginal agreement with the measurements of Higdon
(\cite{Higdon2}) based on HI kinematics. Modeling of the radial $B-V/V-K$
color gradients of the Cartwheel ring galaxy also indicates that the outer
ring is propagating at $v_{\rm ring} \ge 90$~km~s$^{-1}$ for the adopted
distance to the galaxy of 140~Mpc. On the other hand,
the azimuthally averaged $\rm H\alpha$ surface brightness profile of the
Cartwheel's outer ring does not peak exterior to those in {\it K-} and {\it
B}-bands, contrary to what would be expected for such a high propagation
velocity.  We show that a combined effect of 41$^{\circ}$
inclination, finite thickness, and warping of the Cartwheel's disk might
be responsible for the lack of angular difference in the peak positions. Indeed,
the radial $\rm H\alpha$ surface brightness profiles obtained along the
Cartwheel's major axis, where effects of inclination and finite thickness
are minimized, do peak exterior to those at {\it K-} and {\it B}-bands. The
angular difference in peak positions implies $v_{\rm ring}$~=~110~km~s$^{-1}$,
which is in agreement with the model
predictions. We briefly discuss the utility of radio continuum emission and
spectral line equivalent widths for determining the propagation velocity of
gas rings in collisional ring galaxies.
}

\titlerunning{Propagation velocities of gas rings}
\maketitle

\section{Introduction}

Collisional ring galaxies are thought to originate in galactic collisions.
A near-central axial passage of a companion galaxy through the disk of a
target galaxy initiates a series of expanding ring density waves both in
gas and stellar components of the target galactic disk. An expanding
ring gas density wave (hereafter, gas ring) triggers high rates of
massive star formation along its perimeter, if the gas density exceeds a
threshold (Appleton \& Struck-Marcell \cite{Appleton}).  The propagation
velocity of gas ring, i.e. the velocity at which the maximum in radial gas
density profile propagates radially in the galactic disk, is usually inferred from the radial expansion velocity of gas
in the ring. Indeed, kinematical models predict that a determination of the
radial expansion velocity of an individual star or HII region in the first ring
gives a good indication of the ring propagation velocity (Toomre
\cite{Toomre}). This is usually not true for higher order rings.
However, rings are rather wave phenomena than material features, which
implies a possible difference between the ring propagation velocity and
the radial expansion velocity of particles forming the ring.
Indeed, recent N-body modeling of ring galaxy formation by Athanassoula et al.
(\cite{Athan}) have shown that the propagation velocity of the first stellar
ring is usually higher than the mean velocity of stellar particles in the
ring. Hence, the propagation velocity of the first gas ring may also not
coincide with the radial expansion velocity of gas in the ring.  If so, this would
have important implications for the age estimates of collisional ring
galaxies (hereafter, ring galaxies), since in many well-known systems,
most notably in the Cartwheel, the rings are essentially a gas dynamical
phenomenon.

In this paper we attempt to estimate the propagation velocity of gas ring
from the optical and near-infrared photometry, as well as non-thermal radio
emission and spectral line equivalent widths of ring galaxies. If an
expanding gas ring indeed triggers high rates of star formation as predicted
by Appleton \& Struck-Marcell (\cite{Appleton}), then its propagation
velocity might be inferred from the properties of young stellar component
formed in the ring. In Sect.~\ref{gradients} we explore the utility of the
radial optical and near-infrared color gradients measured in the Cartwheel's
disk (Marcum et al. \cite{Marcum}) for estimating the propagation velocity
of the Cartwheel's outer ring. In Sect.~\ref{simulate} we determine the
relationship between the propagation velocity of the first gas ring and
the maximum radial
expansion velocity of gas in the ring.  The Cartwheel's radial surface
brightness profiles in {\it Ks}-band, {\it B}-band and $\rm H\alpha$ are
compared with those predicted by a density wave model in
Sect.~\ref{surf_br}. In Sect.~\ref{other} we discuss the utility of radio
continuum emission and spectral line equivalent widths for determining the
propagation velocity of gas rings in ring galaxies. Our main results are
summarized in Sect.~\ref{sum}.

\section{Radial optical and near-infrared color gradients in the Cartwheel's
disk}
\label{gradients}

Theory of ring galaxy formation predicts that in gas-rich galactic disks an
expanding ring gas density wave might trigger a co-existing expanding
wave of star formation (Appleton \& Struck-Marcell \cite{Appleton}). As such
a density-wave-induced ring wave of star formation propagates radially from
the nucleus, it leaves behind evolved stellar populations, with the youngest
stars located at the current position of the wave. This might result in a
radial color distribution of stars in the galactic disk, with the inner
regions being redder than the outer parts of the disk. Indeed, observations
of the Cartwheel ring galaxy (Marcum et al. \cite{Marcum}) and some other
ring galaxies (Appleton \& Marston \cite{Marston}) revealed the presence of
the optical and near-infrared radial color gradients in the galactic disks.

Recent attempts by Korchagin et al. (\cite{Korch2}) and Vorobyov \& Bizyaev
(\cite{VB}, hereafter VB) to model the Cartwheel's radial $B-V/V-K$ color gradients using
the above scenario yielded a clear success.  One of the results of Korchagin
et al. (\cite{Korch2}) was to show that the model radial $B-V/V-K$ color
gradients are sensitive not only to the metallicity of star-forming gas, but
also to the propagation velocity of gas ring. Indeed, faster rings are
expected to produce younger stellar populations provided that the rings have
propagated to the same diameter in the galactic disk. This motivated us to
investigate the utility of the Cartwheel's radial color gradients for
estimating the ring propagation velocity.

In our previous paper (VB) we postulated the
propagation velocity of the Cartwheel's outer ring as an average of all the
available measurements (55~km~s$^{-1}$) and varied the metallicity of
star-forming gas in the Cartwheel's disk in order to find a better agreement
between the model and observed radial color gradients. In this paper we perform
a more comprehensive modeling of the Cartwheel's radial $B-V/V-K$ color
gradients and $Q_{\rm BVK}$ color indices by varying the propagation
velocity as well. Our intuitive expectations are that the Cartwheel's radial
color gradients cannot be reproduced equally well for any value of the outer
ring propagation velocity. If so, we would be able to estimate the outer
ring propagation velocity from the available optical and near-infrared
photometry of the Cartwheel ring galaxy.  The equations and methods used to
model the colors and color indices are explained in detail in VB and references therein.
The population synthesis code used in this work is explained in detail in
Mayya (\cite{Mayya}, \cite{Mayya2}). Using stellar evolutionary (Schaller et al.
\cite{Schaller}) and atmospheric (Kurucz \cite{Kurucz}) models, this code synthesizes
a number of observable quantities in the optical and near-infrared passbands,
which are suitable for comparison with the observed properties of
giant star-forming complexes. Since the Cartwheel is a recent
phenomenon, this code is especially suitable for our purposes. The
results of this code are compared with those of other existing
codes by Charlot (\cite{Charlot}).

A major difference of our
modeling from that of VB is in the adopted
distance to the galaxy, with the value of a Hubble constant being $H_{\rm 0}$ =
65~km~s$^{-1}$~Mpc$^{-1}$ instead of 100~km~s$^{-1}$~Mpc$^{-1}$.
The need for this change comes from the realization that the results of
our modeling depend on the adopted distance to the galaxy.
For example, choice of $H_{\rm 0}=50$~km~s$^{-1}$~Mpc$^{-1}$ instead
of $H_{\rm 0}=100$~km~s$^{-1}$~Mpc$^{-1}$ would double the linear diameter
of the Cartwheel's disk. The time it takes for the density wave of constant
propagation velocity to cross the Cartwheel's disk would grow accordingly,
which will inevitably influence the colors of density-wave-born stellar
populations in each radial annulus of the Cartwheel's disk.
There are distinct signs of convergence between different estimates of a
Hubble constant to the value of $H_{\rm 0}=65\pm 10$~km~s$^{-1}$~Mpc$^{-1}$
(Binney \& Merrifield \cite{Binney}). Therefore, in this work we adopt
$H_{\rm 0}=65$~km~s$^{-1}$~Mpc$^{-1}$, which assumes 140~Mpc for the
distance to the Cartwheel galaxy.

We consider several propagation velocities of the outer ring $(v_{\rm ring})$ 
starting from 25~km~s$^{-1}$ and ending with 120~km~s$^{-1}$. 
We focus  on three values, namely 25~km~s$^{-1}$,
55~km~s$^{-1}$, and 90~km~s$^{-1}$ as implied by the
measurements of gas expansion velocity by Amram et al. (\cite{Amram}),
Higdon (\cite{Higdon2}), and  Fosbury \& Hawarden (\cite{FH}).
For each value of propagation velocity we explore a wide range of
metallicities from $z=z_{\odot}/20$ to $z=2 z_{\odot}$ in order to achieve a
better agreement between the model and observed photometry.
The properties of the pre-collision stellar disk are chosen
typical for the late-type Freeman disks, with the {\it V}-band central
surface brightness $\mu_{\rm V}^0=21.0~mag~arcsec^{-2}$ and the scale length
$R_0=3$~kpc respectively.  In this paper we do not model the colors of the
Cartwheel's inner ring and the nucleus, as it requires additional
assumptions on the star formation history in the Cartwheel (see VB for
details).

\begin{figure}
  \resizebox{\hsize}{!}{\includegraphics{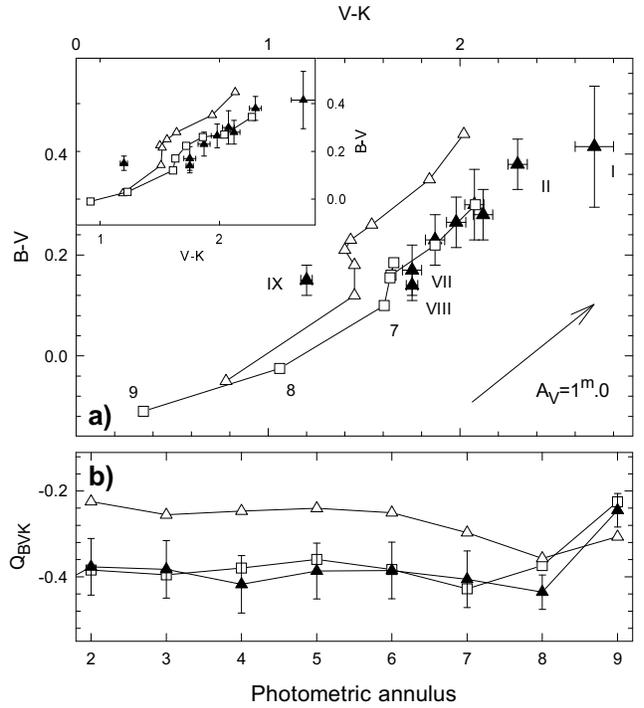}}
      \caption{The radial profiles of {\bf a)} $B-V/V-K$ color gradients
      and {\bf b)} the $Q_{\rm BVK}$ combined color indices. The open squares
      and the open triangles show the model profiles obtained for the outer
      ring propagation velocity
      of 90~km~s$^{-1}$ and 25~km~s$^{-1}$ respectively.
      The filled triangles with error bars indicate {\bf a)} the measured
      $B-V/V-K$ radial color gradients in the Cartwheel's disk and {\bf b)}
      the $Q_{\rm BVK}$ color indices calculated from the measured $B-V/V-K$
      colors of the Cartwheel's disk. The insert shows the results of color
      modeling with Starburst99 (Leitherer et al. \cite{Leitherer}).  }
         \label{Fig1}
   \end{figure}

The Cartwheel's and the model colors/indices obtained for the
outer ring propagation velocity of 25~km~s$^{-1}$ and
90~km~s$^{-1}$ are shown in Fig.~\ref{Fig1}. Open triangles in
Figs.~\ref{Fig1}a and \ref{Fig1}b represent the model radial $B-V/V-K$ color
gradients and $Q_{\rm BVK}$ combined color indices obtained  for
$v_{\rm ring}=25$~km~s$^{-1}$, while the open squares indicate the
corresponding values obtained for $v_{\rm ring}=90$~km~s$^{-1}$.
For the latter velocity, the model colors of three outermost annuli 
are indicated in Fig.~\ref{Fig1}a by Arabic numbers.
The filled triangles with error bars represent the radial $B-V/V-K$ color
gradients and $Q_{\rm BVK}$ combined color indices of the
Cartwheel ring galaxy. The Cartwheel's colors were obtained by Marcum et al.
(\cite{Marcum}) for nine annuli and the nucleus (see their Plate I) and
numbered in Fig.~\ref{Fig1}a according to radius, beginning with the inner
ring (annulus I) and ending with the outer ring (annuli VIII-IX).

The fast outer ring can nicely reproduce the Cartwheel's $Q_{\rm BVK}$ indices,
if metallicity gradient of star-forming gas
ranges from $z=z_{\odot}$ in the inner parts (annuli II-IV) to
$z=z_{\odot}/3.7$ in the outer parts (annuli V-IX).
The $Q_{\rm BVK}$ indices minimize the uncertainties in the observed
$B-V/V-K$ colors introduced by dust extinction. Hence, a good correspondence
between the model and the Cartwheel's $Q_{\rm BVK}$ indices 
indicates that the difference in each annulus between the model $B-V/V-K$
colors shown in Fig.~\ref{Fig1}a by the open squares  and the Cartwheel's $B-V/V-K$
colors might be due to dust extinction
in the Cartwheel's disk. The value of color excess $E(B-V)$ implied by this
difference can be used to estimate the surface densities of gas in each
annulus using standard gas-to-dust ratio
(Bohlin et al. \cite{Bohlin}). The
estimated surface densities of gas are below the detection upper limits for
each annulus, which indicates that internal extinction is indeed responsible
for the difference between the Cartwheel's $B-V/V-K$ colors and the model colors
shown in Fig.~\ref{Fig1}a by the open squares.

On the other hand, the slow outer ring cannot reproduce the Cartwheel's
$Q_{\rm BVK}$ indices. Disagreement between the model and Cartwheel's indices
is less severe, if  metallicity gradient of star-forming gas
ranges from $z=z_{\odot}$ in the inner parts (annuli II-IV) to
$z=z_{\odot}/3.7$ in the outer parts (annuli V-IX).
Nevertheless, the model $Q_{\rm BVK}$ indices shown by the open triangles in
Fig.~\ref{Fig1}b lie beyond the observed limits, which indicates that the
difference in each annulus between the model $B-V/V-K$ colors
shown in Fig.~\ref{Fig1}a by the open triangles and the Cartwheel's colors
cannot be attributed to dust extinction in the Cartwheel's
disk.

We considered other propagation velocities of the outer ring and found that
the Cartwheel's $B-V/V-K$ colors and $Q_{\rm BVK}$ color indices could be
acceptably reproduced for  55~km~s$^{-1}~\le~v_{\rm ring}~\le~$120~km~s$^{-1}$. 
The best-fit value is $v_{\rm ring}=90$~km~s$^{-1}$. 
Stellar populations born in each
annulus by a slower wave are older than those born by a
faster wave. Strong dependence of $B-V/V-K$ colors on the age of young
stellar populations allows us to constrain the
propagation velocity of the Cartwheel's outer ring.

Three assumptions can possibly influence our color modeling results:
the assumed properties of the pre-collision stellar disk,
the adopted value of a Hubble constant, and the population synthesis model
used in color gradient modeling.

{\it a) The pre-collision stellar populations.}
To understand how the model $B-V/V-K$ colors depend on the properties of the
pre-collision disk, we gradually reduced $\mu_{\rm V}^0$ down to 23.5~mag
and increased $R_0$ up to 5~kpc imitating transition from the late-type
galactic disk to the low surface brightness (LSB) disk.  Again, for $v_{\rm
ring}
<50$~km~s$^{-1}$ the model colors and indices totally disagree with those of
the Cartwheel galaxy. The acceptable agreement between the model and the
Cartwheel's radial color gradients and color indices is found for higher
propagation velocities of 90~km~s$^{-1}$, though the model $Q_{\rm BVK}$
indices tend to lie systematically above the observed values. This again
favors to the pre-collision Cartwheel being a normal late-type galaxy
rather than a LSB galaxy (VB).

{\it b) Distance to the Cartwheel}. We recomputed the model colors
for a Hubble constant of $H_{\rm 0}=100$~km~s$^{-1}$~Mpc$^{-1}$,
which assumed 90~Mpc for the distance to the Cartwheel galaxy.
The adopted distance to the Cartwheel affects the model radial color
gradients. A general tendency is that the acceptable agreement between
the model and the Cartwheel's $B-V/V-K$ colors and $Q_{\rm BVK}$ indices
can now be achieved for lower values of the ring propagation velocity, more
specifically, for $v_{\rm ring} \ge 40$~km~s$^{-1}$.  The best-fit
value is now $v_{\rm ring}=60$~km~s$^{-1}$. For the ring propagation velocities
higher than 90~km~s$^{-1}$, the correspondence between the model
and the Cartwheel's colors/indices of the four outermost annuli
worsens.

{\it c) Population synthesis model}. We tested our results of color 
gradient modeling with another population synthesis
code, namely Starburst99 (Leitherer et al. \cite{Leitherer}). This code was developed for synthesizing
observable quantities in starburst galaxies. Since the Cartwheel
harbors a moderate burst of star formation ($SFR \approx 67~M_{\odot}$~yr$^{-1}$),
Starburst99 is best suited for comparison with the code assumed in this
work ( Mayya \cite{Mayya}).
The results of color modeling with Starburst99 are plotted in 
the insert of Fig.~\ref{Fig1}a for both the fast outer ring 
($v_{\rm ring}=90$~km~s$^{-1}$, the open squares) and the slow outer ring
($v_{\rm ring}=25$~km~s$^{-1}$, the open triangles). 
It is clearly evident that  the fast outer ring can well reproduce
the Cartwheel's colors. 
On the contrary, the slow outer ring cannot reproduce the Cartwheel's
optical and near-infrared colors. 
Starburst99 uses a new atmospheric model of Lejeune et al. (\cite{Lejeune})
and the latest Geneva evolutionary models, which are different from those
used by Mayya in his population synthesis code (Mayya \cite{Mayya}). 
Nevertheless, the results of color modeling with both codes are very similar.
Hence, our main conclusions are not affected
by the population synthesis model assumed in this work.

Direct estimate of the Cartwheel's outer ring propagation velocity from the
color gradient modeling performed in this section is problematic since
our results are sensitive to the assumed value of a Hubble constant.
Nevertheless, our  modeling indicates that the
propagation velocity of the Cartwheel's outer ring cannot be lower
than 40~km~s$^{-1}$ for any reasonable value of a Hubble constant.
Determining the ring propagation velocity may be in fact even more difficult, 
because it may not coincide with 
the expansion velocity of gas in the ring. Indeed, rings are rather wave phenomena
than material features (Athanassoula et al. \cite{Athan}) and the
difference between the ring propagation velocity and
the radial expansion velocity of particles forming the ring
is not totally unexpected.
In the next section two dimensional numerical
simulations of a Cartwheel-like ring galaxy are performed. We restrict
ourselves to the planar geometry in order to find the relationship between
the maximum radial expansion velocity of gas in the ring and the
propagation velocity of the gas ring itself.

\section{Numerical simulations of a Cartwheel-like ring galaxy}
\label{simulate}
\subsection{Initial conditions}
\label{init}

An extensive study of the Cartwheel galaxy by Higdon (\cite{Higdon1},
\cite{Higdon2}) provided important observational constrains on the
parameters of our numerical model. The indicative mass of the Cartwheel,
which generally represents the total mass to within a factor of about 2, is
$5.3\times 10^{11}~M_{\odot}$ for the adopted
$H_0=65$~km~s$^{-1}$~Mpc$^{-1}$ (Higdon \cite{Higdon2}).  An estimate of the
luminous stellar mass $M_{\rm st}$ in the Cartwheel is derived by
multiplying its blue luminosity $L_{\rm B}=3.7\times 10^{10}~L_{\odot}$ (Higdon
\cite{Higdon1}) by a late spiral ratio $M/L_{\rm B}$=3 (Faber \& Gallagher
\cite{Faber}). This results in $M_{\rm st}=1.1 \times 10^{11}$~$M_{\odot}$.
The total HI mass of the Cartwheel is $2.2 \times 10^{10}~M_{\odot}$ (Higdon
\cite{Higdon2}, corrected for $H_0=65$~km~s$^{-1}$~Mpc$^{-1}$), with the
total gas mass ($M_{\rm HI}+M_{\rm He}$) amounting to $3 \times
10^{10}~M_{\odot}$. The Cartwheel galaxy thus appears to be halo-dominated.
This assumption was also used by Appleton \& Higdon ({\cite{Appleton2}) in
their hydrodynamic models of the Cartwheel galaxy.

We consider a Cartwheel-like galaxy consisting of a self-gravitating thin
gaseous disk of $M_{\rm d}=3 \times 10^{10}~M_{\odot}$ and a rigid halo of
$M_{\rm h}=5 \times 10^{11}~M_{\odot}$. Marcum et al. (\cite{Marcum}) found
no evidence for old pre-collision stellar populations in the outer parts of
the Cartwheel. Numerical modeling of the radial $B-V/V-K$ color gradients
confirmed that the pre-collision stellar populations dominate only in the
inner parts of the Cartwheel (VB). Moreover, our
numerical simulations indicate that stellar density waves propagate only 2-3
disk scale lengths and become significantly less defined as compared to
gas density waves, the result also found in numerical simulations by
Hernquist \& Weil (\cite{HW}). Hence, we expect that the stellar density
wave would have little effect on the dynamics of gas at the radii of the
Cartwheel's outer ring.

The radial distribution of gas in the pre-collision Cartwheel is unknown.
Our modeling of radial color gradients measured in the Cartwheel's disk
(VB) suggested that this distribution most
probably showed a mild exponential decline with a surface density contrast
between the nucleus and the position of the outer ring being a factor of
3.5. More importantly, the lack of old stellar component at the outer parts
of the Cartwheel implies that the surface density of gas in the
pre-collision Cartwheel was below the critical surface density for
gravitational instability everywhere except probably the innermost radii.
The critical surface density is defined as

\begin{equation}
\Sigma_{\rm crit}=  0.7 {v_{\rm s} k \over \pi G} ,
\label{crit}
\end{equation}
where $v_{\rm s}$ is the speed of sound and $k$ is the epicyclic frequency
(Martin \& Kennicutt \cite{Martin}).

Initially, the rotating gaseous disk  is balanced by self-gravity of the
disk, the radial pressure gradient, and gravity of the spherically symmetric
rigid halo
\begin{equation}
{v_{\rm cir}^2 \over r}={1 \over \Sigma_{\rm g}} {d P \over dr} + {d \over dr}(\Phi_{\rm
d} +\Phi_{\rm h}),
\label{rot_vel}
\end{equation}
where $v_{\rm cir}$ is the rotational velocity, $P$ is the gas pressure,
$\Phi_{\rm h}$ is the halo potential, and $\Phi_{\rm d}$ is the potential of gaseous
disk defined in the polar coordinates as
\begin{eqnarray}
\Phi_{\rm d}(r, \phi) &=& - G~\int_{R_{\rm in}}^{R_{\rm out}}{r^\prime}~d{r^\prime}\times\nonumber
\\
& & \int_0^{2\pi} {\Sigma_{\rm g}({r^\prime
}, {\phi^\prime})~d{\phi^\prime} \over \sqrt {r^2
+ r^{\prime 2} - 2rr^{\prime}~cos(\phi - \phi^\prime)}}.
\label{disk_pot}
\end{eqnarray}

We adopt an isothermal equation of state ($T=10^4~K^\circ$), which is an usual approximation
used for modeling the interstellar medium in disk galaxies (Mihos \&
Hernquist \cite{Mihos}). Struck (\cite{Struck}) has demonstrated that the
ring structure and dynamics in the isothermal and non-isothermal simulations
are similar. We also tried an adiabatic equation of state and found little
difference in the first ring structure and dynamics, though the second ring
structure and dynamics showed substantial differences with respect to the
isothermal case.

A slightly modified form of rotation curve in a softening point-mass potential
is adopted
\begin{equation}
v_{\rm cir}(r)={v_{\rm 0}~r^{0.9} \over (r^2 +R_{\rm 0}^2)^{0.5} },
\label{rot_curve}
\end{equation}
where $R_{\rm 0}$ is the effective halo radius. The power index 0.9 allows
us to model a Keplerian decline in rotation curves at larger radii.
Parameters of the rotation curve (eq. \ref{rot_curve}) are chosen so that
the halo mass
\begin{equation}
M_{\rm h}={R^2_{\rm out} \over G} {d \Phi_{\rm h} \over dr} \Big|_{r=R_{\rm out}}
\end{equation}
is equal to $5 \times 10^{11}~M_{\odot}$.  As the rotational curve is fixed
by eq. (\ref{rot_curve}) and the disk potential $\Phi_{\rm d}$ is determined
by numerically integrating eq. (\ref{disk_pot}), the gradient of halo
potential $\Phi_{\rm h}$ can be calculated using eq. (\ref{rot_vel}).
Parameters used in numerical simulations are listed in Table~\ref{Table1}.

\begin{table}[h]
\caption{Model parameters of a Cartwheel-like galaxy}
\vskip 0.1 cm
\begin{tabular}{ll}
\hline
\hline
Parameter &  Value \\
\hline
Inner disk radius ($R_{\rm in}$) & 0.2 kpc  \\
Outer disk radius ($R_{\rm out}$) & 34 kpc \\
Disk scale length ($r_{\rm 0}$) & 20 kpc \\
Central surface density of gas ($\Sigma_{\rm g}(0))$ & 23.5~$M_{\odot}$~$pc^{-2}$
\\
Gas temperature ($T$) & $10^4~ {\rm K}^\circ$ \\
Disk mass ($M_{\rm d}$) & $3 \times 10^{10}~M_{\odot}$ \\
Halo mass ($M_{\rm h}$) & $5 \times 10^{11}~M_{\odot}$ \\
Mass of a companion ($M_{\rm c}$) & $1.5 \times 10^{11}~M_{\odot}$ \\
Companion's softening parameter ($\epsilon$) & 5~kpc \\
\hline
\end{tabular}
\label{Table1}
\end{table}

\subsection{Computational techniques}

We use the ZEUS-2D numerical hydrodynamics code incorporating a
self-gravitating thin gaseous disk and a fixed spherically symmetric halo
potential. A usual set of isothermal hydrodynamics equations in the polar
coordinates ($r, \phi$) is solved using the method of finite-differences
with a time-explicit, operator split solution procedure explained in detail
in Stone \& Norman (\cite{Stone}). The interaction is treated by computing
the gravitational force exerted on the disk by the softened point-mass
potential of a companion galaxy. Equations of motion of the companion in the
combined gravitational potential of the halo and the gaseous disk are
numerically solved using the Runge-Kutta scheme. The gravitational potential
of the thin disk (eq. \ref{disk_pot}) is evaluated numerically using the
fast Fourier transform (Binney
\& Tremaine 1987). The code was verified against a standard set of test
problems described in Stone \& Norman (\cite{Stone}).

\subsection{Results of modeling.}
We attempt to find the relationship between the propagation velocity of ring
gas density wave and the maximum radial expansion velocity of gas in
the wave generated by a near central passage of a companion galaxy through the disk of a
Cartwheel-like galaxy. We have chosen the model parameters typical for the
Cartwheel galaxy in order to compare the results of numerical simulations
with those of Sect.~\ref{gradients}. By varying the initial rotation curve,
the mass of a companion, and the impact point we try to reproduce the morphology, kinematics, and radial
distribution of gas in the Cartwheel's disk. 
Below we briefly describe the parameter space investigated in our simulations.

{\it Initial rotation curve}. We consider three initial rotation
curves shown in Fig.~\ref{Fig4} and find that the
Cartwheel's HI morphology and kinematics are best reproduced for A2 curve. 
A1 curve produce too small a spacing between the inner and outer rings as 
compared to that of the Cartwheel, while A3 curve does not develop the inner ring 
at all.

{\it Mass of a companion}. We consider a wide range of companion galaxy masses 
from $4 \times 10^{10}~M_{\odot}$ to $3 \times 10^{11}~M_{\odot}$, which 
is roughly $7.5\% - 55\%$ of the Cartwheel's total mass.  
The indicative mass of a companion G3, most massive in the Cartwheel's group
(Higdon \cite{Higdon2}), is $M_{\rm G3}= 4 \times 10^{10}~M_{\odot}$. This
mass is not enough to reproduce the radial surface density profiles of HI in
the Cartwheel. Such a low-mass companion gives a weak first response, with
less than a $20\%$ enhancement in gas densities over the unperturbed values.
At the time when the first ring reaches 24~kpc radius (the
current location of the Cartwheel's outer ring), the actual mass of gas
locked in the ring is only $25\%$ of the total gas mass, while in the
Cartwheel the outer ring contains about 85\% of total HI (Higdon
\cite{Higdon2}). The low mass companion does not reproduce the observed
spacing between the inner and the outer Cartwheel's rings, with the radius
of the model second ring being almost twice bigger than that of the
Cartwheel's inner ring.

\begin{figure}
  \resizebox{\hsize}{!}{\includegraphics{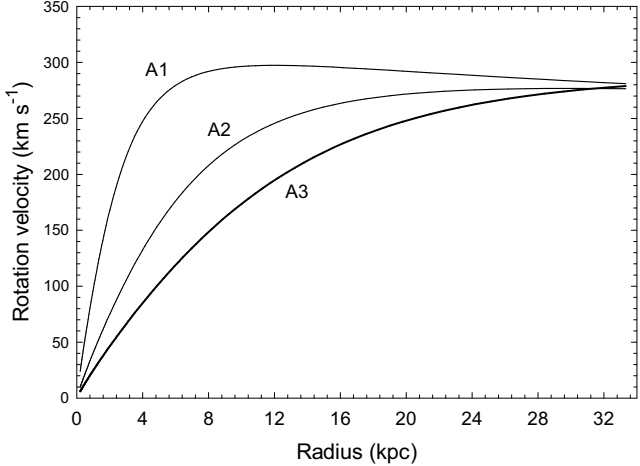}}
      \caption{Rotation curves of pre-collision gaseous disk generated
      using eq.~\ref{rot_curve}. Parameters determining rotation curves
      are: $v_0=375$ and $R_0=4$ (marked "A1"), $v_0=382$ and $R_0=10$
      (marked "A2"), $v_0=420$
      and $R_0=18$ (marked "A3").
      The Cartwheel's outer ring is at 24~kpc ($H_0=65$~km~s$^{-1}$~Mpc$^{-1}$).
              }
         \label{Fig4}
   \end{figure}

We find that companions with masses no less than $25\%$ of the Cartwheel's 
total mass can acceptably reproduce the HI distribution in the Cartwheel's 
disk. In our simulations the companion's mass is  $1.5 \times 10^{11}~M_{\odot}$, 
which is roughly $30\%$ of the total mass of the Cartwheel. A $30\%$ companion 
generates stronger response, with more than a 100\% enhancement in gas surface 
densities over the unperturbed values.
There are no companions to the Cartwheel large enough to fit the companion's mass 
we determine. The companion's optical and HI line widths give mass estimates 
that are only $\sim 6\%$ of the Cartwheel's (Davies \& Morton {\cite{Davies}, 
Higdon \cite{Higdon2}). However, determining accurate galaxy masses is 
difficult enough in strongly perturbed galaxies like the Cartwheel's companions. 
Passing through the Cartwheel's disk causes the companion to lose mass due to tidal 
and ram pressure stripping. Indeed, Higdon (\cite{Higdon2}) reported HI plumes 
stretching from the Cartwheel toward G3 companion. Ring galaxy simulations 
by Tsuchiya et al. (\cite{Tsuchiya}) found that massive companions 
($25\%$ of ring galaxy's mass) can loose around $35\%$ of their halos, or nearly 
a third of their total mass, due to the interaction. The gas disks in massive 
companions were also greatly truncated ($R_{\rm gas} < 50\%$ of original radius). 
If much of the companion's mass resided in a halo, one would in fact significantly 
underestimate its total mass. The apparent low masses of the companion galaxies 
are a problem, but it is likely to be a consequence of the ring galaxy formation 
process. More work is clearly needed in estimating masses of ring galaxy 
companions to see if this is a general result.

{\it Impact points, impact angles, and relative velocities.} 
In our simulations the companion placed initially at a distance 
of 35~kpc above the Cartwheel's plane, which was roughly $150\%$ 
of the Cartwheel's radius, was given
an initial relative velocity of 80~km~s$^{-1}$ toward the Cartwheel.    
Varying the initial distances from the axis of the Cartwheel,
we obtained a wide range of impact points starting from a nearly bull's-eye 
collision ($R_{\rm impact}$=0.1~kpc) 
and ending with an essentially off-center collision ($R_{\rm impact}$=10~kpc). 
The former produces symmetric rings with negligible azimuthal variations in gas 
surface densities around the ring, while the latter results in crescent 
or spiral-like density waves. Neither of these ultimate cases resembles 
the actual HI distribution in the Cartwheel's outer ring. We find 
that the Cartwheel's HI kinematics and morphology are best reproduced 
for a slightly off-center $R_{\rm impact}$=2.5~kpc collision. The relative velocity
of the companion at the time of impact is 480~km~s$^{-1}$.
We also considered other distances above the Cartwheel's plane and
other relative velocities as the initial parameters of the
companion. For the initial distance of 70~kpc and initial relative velocity of 
80~km~s$^{-1}$ toward the Cartwheel, the relative velocity at 
the time of impact was $v_{\rm relative}(impact)=540$~km~s$^{-1}$. 
For the initial distance of 35~kpc and initial relative velocity of 
380~km~s$^{-1}$ toward the Cartwheel,  $v_{\rm relative}(impact)$ was
610~km~s$^{-1}$. Hence, $v_{\rm relative}(impact)$ is largely determined 
by the parameters of the Cartwheel. Variations of $15\%$ in the relative velocity
affect insignificantly the gas dynamics after the collision. 
Angles of impact from the axis of the Cartwheel are 5-15$^{\circ}$  in all considered cases. 
We have not made special efforts to model highly inclined collisions. 

\begin{figure}
\bigskip
\centerline{See figure3.gif}
\bigskip
      \caption{The model velocity field of gas superimposed on the contour plot
of gas surface density $\Sigma_{\rm g}$ obtained
for A2 curve  and model parameters listed in Table~\ref{Table1}. The solid line outlines
the regions where $\Sigma_{\rm g}$ exceeds the critical surface density
for gravitational instability defined by eq. \ref{crit}. The scale bar
is in $\rm M_{\odot}$~pc$^{-2}$.
      }
         \label{Fig5}
   \end{figure}

Figure~\ref{Fig5} shows the model velocity
field of gas superimposed on the contour plot of gas surface density
$\Sigma_{\rm g}$ obtained for A2 curve and model parameters listed in
Table~\ref{Table1}. The contrast between $\Sigma_{\rm g}$ in the outer ring
and the inter-ring region is in agreement with observations. $\Sigma_{\rm
g}$ between the rings in Fig.~\ref{Fig5} never exceed 7~$\rm M_{\odot}$~pc$^{-2}$ and falls
below 5~$\rm M_{\odot}$~pc$^{-2}$ behind the densest regions in the outer
ring. Maximum and average $\Sigma_{\rm g}$ around the outer ring are 18.5
and 15~$\rm M_{\odot}$~pc$^{-2}$, however the observed high density
condensations up to 60~$\rm M_{\odot}$~pc$^{-2}$ are not found in
simulations. $\Sigma_{\rm g}$ is below 7~$\rm M_{\odot}$~pc$^{-2}$ beyond
the outer ring. High density clumps are found in the inner ring and central
regions, with $\Sigma_{\rm g} \approx$~60-100~$\rm M_{\odot}$~pc$^{-2}$ and even
higher. Development of such high density clumps is interesting, since high
densities and low temperatures are thought to favor HI~$\rightarrow~H_2$ transitions
in giant gas clouds. Indeed, Horrellou et al. (\cite{Horellou2}) detected a
substantial amount of molecular hydrogen in the Cartwheel's inner part.
For the solar metallicity implied by color gradient modeling in Sect.~\ref{gradients},
$H_2$ mass within the inner ring amounts to $2.1 \times 10^9~M_{\odot}$
according to Horellou et al. (\cite{Horellou2}). In our simulations, the
inner ring contains $\le 10\%$ of the gas mass or $\le 3 \times
10^9~M_{\odot}$, which is roughly consistent with the detected $H_2$ mass.
In Sect.~\ref{init} we neglected a possible contribution of molecular hydrogen to the 
gas mass of the Cartwheel as it was only a small fraction
to the total mass of the Cartwheel galaxy ($\le 1\%$) and would not affect
noticeably the companion's dynamics.

While the inner ring in Fig.~\ref{Fig5} has on average higher surface densities of gas
than the outer ring, it contains less than $10\%$ of the total disk mass.
On the contrary, the outer ring in Fig.~\ref{Fig5} comprises more than
$50\%$ of the total disk mass. No obvious spokes were
obtained for either of three initial rotation curves. Indeed, HI
observations of Higdon (\cite{Higdon2}) have shown that the Cartwheel's
spokes are HI poor. The solid line in Fig.~\ref{Fig5} outlines the regions
where $\Sigma_{\rm g}$ exceeds the critical surface density for
gravitational instability defined by eq. (\ref{crit}). The regions where star
formation is supposed to occur correlate with the regions of maximum
$\Sigma_{\rm g}$ in the outer ring. This is not the case for the Cartwheel,
where a pronounced anticorrelation between $\Sigma_{\rm HI}$ and $\rm H\alpha$
surface brightness was found (Higdon \cite{Higdon2}). Among other reasons such as
a blow-out in giant supershells formed by multiple supernova explosions, this
might be explained as the consumption of the gas supply by the starburst.
Inclusion of star formation processes in the code is required to check
this hypothesis.

\begin{figure}
  \resizebox{\hsize}{!}{\includegraphics{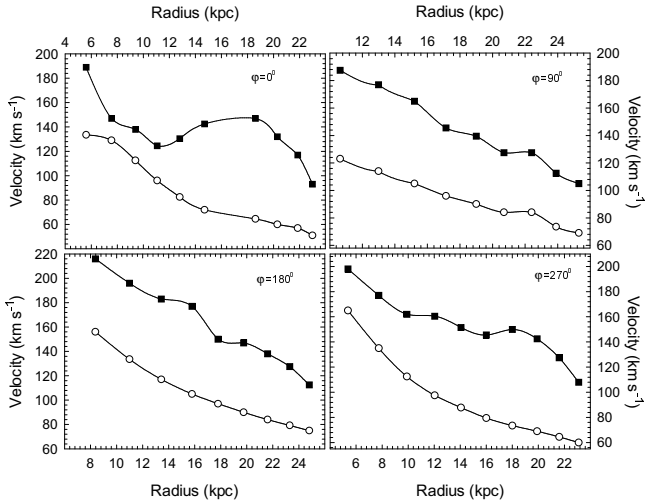}}
      \caption{The propagation velocities of the outer ring (the filled
      squares) and the maximum expansion velocities of gas in the ring (the
      open circles) traced along four different polar angles as indicated
      in each frame. The propagation velocities of the outer ring are
      systematically higher than
      the maximum expansion velocities of gas in the ring.
      }
         \label{Fig6}
   \end{figure}

The velocity field in Fig.~\ref{Fig5} shows all basic features of HI
kinematics found in the Cartwheel: (1) a general increase in expansion
velocity of gas for $R>16$~kpc, (2) inflow of gas towards the inner ring for
$R<16$~kpc. The flow of gas inside the inner ring is complicated. However, a
general expansion of gas is clearly seen in this innermost region.  We
determine the relationship between the propagation velocity of the outer
ring $v_{\rm ring}$ (i.e. the propagation velocity of gas density wave) and
maximum radial expansion velocity of gas $v_{\rm gas}$ in the outer ring at
different ring expansion phases. The regions of maximum density enhancements
in the outer ring are used to trace the outer ring position during the
runs. We start from very early times when the
outer ring was only $\approx 6$~kpc in radius and stop at the outer ring
current position of $\approx 24$~kpc.  Due to obvious asymmetric structure
of $\Sigma_{\rm g}$ in Fig.~\ref{Fig5}, these velocities were traced along
four different polar angles of $0^\circ$, $90^\circ$, $180^\circ$, and
$270^\circ$. The results are shown in Fig.~\ref{Fig6}. The propagation
velocities of the outer ring (the filled squares) are systematically higher
than the maximum radial expansion velocities of gas in the ring (the open circles).
Both velocities generally decline as the outer ring propagates outwards. At
the time when the outer ring reaches its current location at the radius
of 24~kpc, $v_{\rm ring}$ and $v_{\rm gas}$ slow down on average to
102~km~s$^{-1}$ and 65~km~s$^{-1}$ respectively. However, sinusoidal
azimuthal variations in the values of both velocities are evident in
Fig.~\ref{Fig6}.  The value of $v_{\rm ring}=102$~km~s$^{-1}$ is consistent
with the estimates of Sect.~\ref{gradients}, where best agreement between
the the model and observed radial $B-V/V-K$ color gradients was found for
$v_{\rm ring} \ge 90$~km~s$^{-1}$. The value of $v_{\rm gas}=65$~km~s$^{-1}$
is close to the upper estimate of Higdon (\cite{Higdon2}) based on HI
kinematics in the Cartwheel's outer ring. The value of $v_{\rm gas}=13-30
\pm 10$~km~s$^{-1}$ obtained by Amram et al. (\cite{Amram}) using $\rm
H\alpha$ kinematics of ionized gas stands apart. It is known that dynamics
of different interstellar phases may considerably differ in disk galaxies.
Inclusion of multiphase gas dynamics and star formation in the code is
needed to determine the dynamical properties of ionized gas in ring
galaxies.

\begin{figure}
  \resizebox{\hsize}{!}{\includegraphics{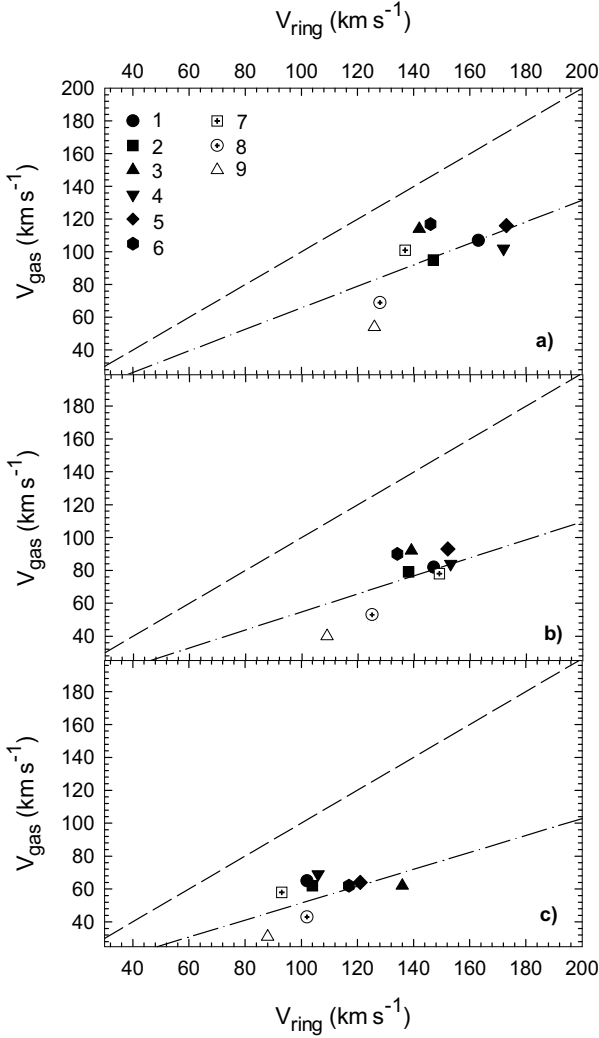}}
      \caption{The maximum expansion velocities of gas in the ring 
as a function of propagation velocities of the first ring
at the radii of {\bf a)}~12~kpc, {\bf b)}~18~kpc, 
and {\bf c)}~22~kpc. Parameters of nine runs labeled by Arabic numbers 
in the upper frame are listed in Table~\ref{Table2}.
The least-squares fit to a linear function $v_{\rm gas}=k~v_{\rm ring}$
is shown in each panel by a dash-dotted line. The coefficient $k$ is defined
in the text. A linear function $v_{\rm gas}=v_{\rm ring}$ 
is plotted for comparison in each panel by a dashed line.}
         \label{Fig7}
   \end{figure}

We performed a series of computations with different values of $M_{\rm d}$,
$M_{\rm h}$, $M_{\rm c}$, $r_{\rm 0}$, and different initial rotation
curves. In Figs.~\ref{Fig7}a-\ref{Fig7}c we plot the maximum expansion velocities
of gas in the ring at the radii of 12 (the upper frame), 18 (the
intermediate frame), and 22~kpc (the lower frame) as a
function of the azimuthally averaged propagation
velocities of the first ring at the same
radii.  Parameters of each run labeled by Arabic numbers in Fig.~\ref{Fig7}a 
are listed in Table~\ref{Table2}. Impact points and impact angles 
vary slightly in the range 1.5-3~kpc and 5-15$^{\circ}$ respectively.  
We find that the inequality $v_{\rm ring} > v_{\rm gas}$
holds in all cases, though the resulted morphology and kinematics of gas may
hardly resemble those of the Cartwheel. Both velocities are generally higher
at the smaller radii and they decline at the larger radii. Higher mass
companions produce a stronger response with higher ring propagation
velocities and maximum radial expansion velocities of gas in the ring, thus leaving the
relation between them unchanged. It appears that the $v_{\rm ring} > v_{\rm
gas}$ inequality holds for the first gas rings triggered by the passage of a
companion through the disk of a halo-dominated target galaxy. 
The least-squares fit to a linear function $v_{\rm gas}=k~v_{\rm ring}$
shown in each panel of Fig.~\ref{Fig7} by a dash-dotted line gives
$k=0.67$ for the rings of 12~kpc, $k=0.55$ for the rings of 18~kpc, and
$k=0.51$ for the rings of 24~kpc. A linear function $v_{\rm gas}=v_{\rm ring}$ 
is plotted for comparison in each panel of Fig.~\ref{Fig7} by a dashed line.
We also considered gas rings in ring galaxies with comparable disk
and halo masses, for which we obtained 
an average value of $k=0.8$ for the rings of 12, 18, and 24~kpc.  
In higher order resonant rings (i.e. the second rings) the
flow is more complicated and $v_{\rm ring} > v_{\rm gas}$ inequality is no
more valid. The $v_{\rm ring} > v_{\rm gas}$ inequality contradicts the
predictions of a simple kinematical model of ring galaxy formation, which
argues that a determination of the expansion velocity of an individual star
or HII region in the first ring gives a good indication of the ring
propagation velocity (Toomre \cite{Toomre}).

Higdon (\cite{Higdon2}) reported the narrow HI line widths and the absence
of strong shear in the Cartwheel's outer ring. He concluded that the outer
ring appeared to initiate star formation activity primarily through concentration
of the interstellar medium in the ring rather than through high-speed disruptive
cloud collisions ($v_{\rm relative} \ge 100$~km~s$^{-1}$) as advocated
by Olson \& Kwan (\cite{Olson}) for starburst activity in interacting/merging
galaxies. Apparent lack of molecular hydrogen in the outer ring 
(Horellou \cite{Horellou}) 
argues against compact dense clouds. Hence, we have chosen a smooth initial gas distribution for
the pre-collision Cartwheel rather than a clumpy medium. In the wider context
of a disparity in ring propagation velocity versus gas expansion velocity
in ring galaxies, the $v_{\rm ring} > v_{\rm gas}$ inequality has to be
proved for a clumpy gas distribution. In this connection, sticky particle 
codes (Combes \& Gerin \cite{Combes}) could be very useful. We note 
here that N-body simulations of stellar rings in ring
galaxies by Athanassoula et al. (\cite{Athan}) have also shown that the
expansion velocity of the first stellar ring is usually higher than the mean
velocity of stellar particles constituting the ring 
(their Fig.~17, the upper frame). Hence, a disparity in ring propagation velocity 
versus expansion velocity of gas/stars constituting the ring  might
be a general property of ring galaxies.

\begin{table}[h]
\caption{Model parameters of nine runs shown in Fig~\ref{Fig7}}
\vskip 0.1 cm
\begin{tabular}{llllllp{0.25cm}p{0.25cm}p{0.25cm}p{0.25cm}}
\hline
\hline
N & $M_{\rm d}^{1}$ & $M_{\rm h}^{1}$  & $M_{\rm c}^1$ & $r_{\rm 0}^2$
&$RC^3$&$v_{\rm ring}^{4}$&$v_{\rm gas}^{4}$
& $v_{\rm ring}^5$  & $v_{\rm gas}^5$      \\
\hline
1 & 3 & 50 & 15 & 20 & A2 &  163 & 107 & 102 & 65 \\
2 & 3 & 50 & 15 & 20 & A1 &  147 & 95 & 104 & 62 \\
3 & 3 & 50 & 15 & 10 & A2 &  142 & 114 & 136 & 62  \\
4 & 3 & 50 & 15 & 100 & A2 &  172 & 102 & 105 & 68 \\
5 & 3 & 50 & 21 & 20 & A2 & 173 & 116  & 121 & 64 \\
6 & 3 & 50 & 15 & 20 & A3 &  146 & 117 & 117 & 62 \\
7 & 6 & 45 & 15 & 20 & A2 &  137 & 101 & 93 & 58 \\
8 & 3 & 50 & 10 & 20 & A2 & 128 & 69 & 102 & 43 \\
9 & 3 & 50 & 7.5& 20 & A2 &  126 & 54 & 88 & 31 \\
\hline
\end{tabular}
\vskip 0.2 cm
$^1$ in $10^{10}~M_{\odot}$. \\
$^2$ in kpc. \\
$^3$ Rotation curve as labeled in Fig.~\ref{Fig4}. \\
$^4$ $v_{\rm ring}$ and $v_{\rm gas}$ at 12~kpc radius. \\
$^5$ $v_{\rm ring}$ and $v_{\rm gas}$ at 22~kpc radius (position of the Cartwheel's outer ring).

\label{Table2}
\end{table}

The $v_{\rm ring} > v_{\rm gas}$ inequality for the first gas ring in ring
galaxies has an important implication. Measurements of the radial expansion
velocity of gas in the Cartwheel's outer ring based on HI and $\rm
H\alpha$  kinematics appear to underestimate the true propagation
velocity of the outer ring. 
Moreover, the radial expansion velocity of gas itself may be underestimated. 
Indeed, $v_{\rm gas}$ is derived assuming that all the gas motions are 
in the plane of the ring.
The plane geometry neglects any non-planar component of $v_{\rm gas}$ that
may be present in warped rings. Results of numerical experiments performed
by Struck (\cite{Struck}) indicate that the Cartwheel's disk might be slightly
warped. Hence, the measurements of gas expansion velocity in the Cartwheel's
outer ring may underestimate $v_{\rm gas}$, which
in turn underestimates the true propagation velocity of the outer ring. 
The same might be true for other ring galaxies.
If so, ring galaxies are generally younger than they have been previously
thought. If rings are younger then we can expect to see even fewer of them~-~they 
should be relatively rarer.

\section{Radial {\it B}-band, {\it K}-band, and $\rm H\alpha$ surface brightness profiles of the
Cartwheel galaxy}
\label{surf_br}

Marston \& Appleton (\cite{Marston2}) found that unlike normal galaxies the
azimuthally averaged radial intensity profiles in $\rm H\alpha$ and 
{\it K}-band of most ring galaxies peak at the position of the ring.
Specifically, for the majority of ring galaxies the peak in the $\rm H\alpha$
radial profile lies at the larger radii than the peak in the {\it K}-band radial
profile. Indeed, a standard
scenario of ring galaxy formation predicts
that the {\it K}-band and $\rm H\alpha$ peaks in radial intensity profiles
of density-wave-born stellar populations are spatially separated, with the
peak in $\rm H\alpha$ lying on the leading edge of the wave where most star
formation takes place, and the peak in {\it K}-band lying progressively
behind the wave as the density-wave-born stellar populations evolve
(Korchagin et al. \cite{Korch}).

\begin{figure}
\bigskip
\centerline{See figure3.gif}
\bigskip
      \caption{The model {\it K}-band (filled squares), {\it B}-band (open circles),
      and $H\rm\alpha$ (filled triangles) radial surface brightnesses of the Cartwheel ring
      galaxy obtained for the same model parameters as in Fig.~\ref{Fig1}.
      The surface brightness profiles peak at/near the position of the
      outer ring at 24~kpc.
      Difference in the peak positions  at {\it K}-band and  $\rm H\alpha$
      is bigger than at {\it B}-band and $\rm H\alpha$.}
         \label{Fig8}
   \end{figure}

Applying the model used to compute radial color gradients in
Sect.\ref{gradients}, we found that the linear difference
between the {\it K}-band and $\rm H\alpha$ peak positions depends not only on
the wave velocity, but also on the metallicity of star-forming gas.
Specifically, this difference is proportional to the wave velocity. For
metallicities of star-forming gas below 1/5 of the solar there is no peak in
{\it K}-band at the position of the ring and the model {\it K}-band
radial surface brightness profile shows a monotonous decline at all radii.
The peak appears at higher metallicities and becomes more pronounced with
increasing metallicity. In Fig.~\ref{Fig8} we present the model azimuthally
averaged radial surface brightness profiles of the Cartwheel galaxy in {\it
K}-band (filled squares), {\it B}-band (open circles), and $\rm H\alpha$
(filled triangles). All model parameters of the gas density wave and
pre-collision stellar populations are chosen as in Fig.~\ref{Fig1}.  The
model surface brightness profiles peak at/near the position of the
Cartwheel's outer ring at 24~kpc.  Analysis of Fig.~\ref{Fig8} indicates
that the difference between the $\rm H\alpha$ and
{\it K}-band  peak positions is expected to be around 1.1~kpc (1.65$^{\prime\prime}$) for the
adopted distance of 140~Mpc and the outer ring propagation velocity of 90~km~s$^{-1}$. The ${\rm
H\alpha}-B$ peak difference is 0.7~kpc (1.05$^{\prime\prime}$).  We note
here that the underlying ring stellar density wave may influence the peak
position in {\it K}-band. To exclude this effect, density-wave-born stellar
populations must dominate the ring emission, which appears to be the case for
the Cartwheel galaxy (Marcum et al. \cite{Marcum}, VB).

We compare our model predictions  with the actual angular differences
between the peaks in the azimuthally averaged {\it B}-band, {\it Ks}-band and $\rm H\alpha$ radial
surface brightness profiles of the Cartwheel's outer ring.  Images of the Cartwheel
together with the calibrating frames were obtained from the public archives.
The near-infrared images in  {\it Ks}-band were provided by ESO VLT
archive. Images in the HST's WFPC2/F450W (B) band
were taken from the Hubble Space Telescope archive.
Hereafter, we refer to this band as the "{\it B}-band". Images in
$\rm H\alpha$ and "$\rm H\alpha$-free continuum" were provided by the
CFHT archive. All the data were reduced in the standard way.  Since we were
interested only in determining the angular difference between the peak
positions in different passbands, the flux calibration was not applied.  All
the images were aligned using the standard MIDAS routines. The estimated
accuracy of alignment is better than 0.03$^{\prime\prime}$. Finally, an
angular resolution of 0.76$^{\prime\prime}$ was achieved for
the ${\rm H}\alpha$ image. When the {\it B} and {\it Ks}
images are considered separately from the $\rm H\alpha$ image, a better
resolution of $0.46^{\prime\prime}$ is achieved.
At the adopted distance of 140~Mpc, this corresponds to a spatial resolution
of 500~pc for the ${\rm H}\alpha$ image and 300~pc for the images in {\it
Ks-} and {\it B-}bands.

\begin{figure}
\bigskip
\centerline{See figure3.gif}
\bigskip
      \caption{ The Cartwheel galaxy in {\it Ks}-band. The dashed ellipses outline
      elliptical annuli used for determining the azimuthally averaged radial surface
      brightnesses of the Cartwheel. The north is shown by
      the arrow.
              }
         \label{Fig9}
   \end{figure}

Figure~\ref{Fig9} shows the image of the Cartwheel galaxy in {\it
Ks}-band. The north is down and the east is to the right. As is seen in
Fig.~\ref{Fig9}, the geometrical center of the outer ring does not coincide
with the nucleus.  To take this peculiar morphology into account, a series
of elliptical annuli were constructed, with the smallest ellipse centered on
the nucleus and the centers of larger ellipses slightly shifted so as the
largest ellipse is centered on the geometrical center of the outer ring.
These elliptical annuli were then used to determine the azimuthally averaged
{\it Ks}-band, {\it B}-band, and $\rm H\alpha$ radial surface brightness
profiles shown in Fig.~\ref{Fig10}.

\begin{figure}
  \resizebox{\hsize}{!}{\includegraphics{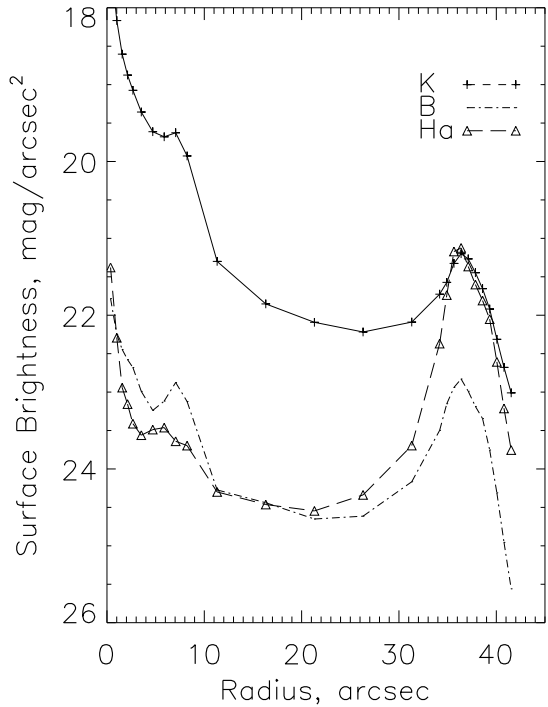}}
      \caption{The azimuthally averaged surface brightness profiles of
      the Cartwheel galaxy. All profiles peak at the position of the outer
      ring at 36$^{\prime\prime}$. No difference between the peak positions
      is seen.}
         \label{Fig10}
   \end{figure}

Surprisingly, we find no noticeable angular difference between the peak
positions of the azimuthally averaged radial surface brightness profiles
in the Cartwheel's outer ring,
contrary to what is expected from our modeling (see Fig.~\ref{Fig8}).
Absence of such a difference is also evident in Fig.~4 of Higdon
(\cite{Higdon1}). This implies that either the outer ring velocity is too
slow for this difference to be resolved or considerable inclination of the
Cartwheel acts to smear out this difference.  Indeed, if the Cartwheel's
disk is warped (Struck \cite{Struck}), then the elliptical annuli might not
be a good choice for obtaining the azimuthally averaged radial profiles.
Another geometry effect may also affect the azimuthally averaged profiles in
considerably inclined ring galaxies. For an inclined disk of finite
thickness, the line-of-sight flux comes from density-wave-born stellar
populations located at different radii, and hence having different ages and
different photometric properties.  This line-of-sight overlapping of stellar
populations of different ages may have a significant effect on the
azimuthally averages surface brightness profiles, if most of the radiation
flux comes from regions close to the minor axis and the line-of-sight
thickness of the disk is comparable to the expected shift between the peaks
in different passbands.

Rough estimates of this effect can be done by assuming that the total
stars+gas surface density in the ring ($\Sigma_{\rm tot}$) is of order 20 $\rm
M_{\odot}$~pc$^{-2}$ and using a typical value of the stellar velocity
dispersion of young stellar populations $c_z=10$~km~s$^{-1}$. For the case
of a self-gravitating disk, the isothermal scale height of the ring $h = c^2
/ 2 \pi G \Sigma_{\rm tot}$ is near 0.2~kpc. The vertical thickness of the outer ring
$2h$ is thus 0.4~kpc.  The gravitational potential of a dark halo makes a
disk thinner, but not very significantly (Bahcall \cite{Bahcall}). At the
same time, disk thickness is expected to be twice bigger for interacting
galaxies (Reshetnikov \& Combes \cite{Resh}). Results of numerical
experiments also indicate that the Cartwheel's outer ring might be thick and
slightly warped (Struck \cite{Struck}).

Our model predicts that the linear difference between the $\rm H\alpha$
and {\it K}-band peaks is
of order of a kiloparsec and this value is thus comparable to the
line-of-sight thickness of the Cartwheel outer ring for the adopted
inclination of $41^{\circ}$ (Higdon \cite{Higdon2}). Hence, we expect that
both the line-of-sight overlapping and the warping act to smear out the
difference between the peak positions of the azimuthally averaged radial
surface brightness profiles shown in Fig.~\ref{Fig10}. Fortunately,
overlapping should have little effect in the regions located along the major
axis of the Cartwheel.

\begin{figure}
  \resizebox{\hsize}{!}{\includegraphics{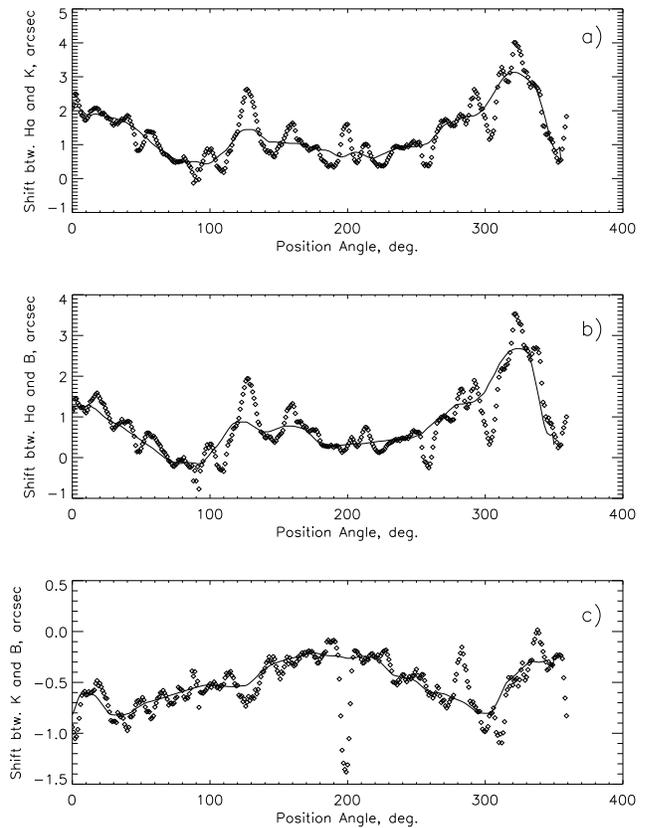}}
      \caption{The angular difference between the peaks in radial
      surface brightnesses azimuthally averaged over a 1$^{\circ}$ sector as
      a function of position angle measured counterclockwise from a line
      extending north from the outer ring center. {\bf a)} Angular
      difference between the $\rm H\alpha$ and {\it Ks}-band peaks, 
      {\bf b)} angular difference between the $\rm H\alpha$ and {\it B}-band
      peaks, {\bf c)} angular
      difference between the {\it Ks-} and {\it B}-band peaks. 
      Solid lines indicate the
      angular peak differences smoothed by a wide median filter ($30^\circ$). }
         \label{Fig11}
   \end{figure}

We apply an alternative method for determining the angular difference
between the peak positions of radial profiles in different photometric bands. Instead of
computing the fluxes from each elliptical annulus shown
in Fig.~\ref{Fig9}, i.e. the fluxes averaged over the whole azimuth,
we make a series of radial cuts drawn from the
Cartwheel's nucleus with an angular separation of $0.2^{\circ}$.  A series
of 360 radial surface brightness profiles of the Cartwheel's outer ring were
then constructed by azimuthally averaging these cuts over a $1^{\circ}$
sector.  Peaks were located by both fitting a Gaussian to each profile and
determining the "center of mass" of each profile. The radial positions of
each peak were corrected for a 41$^{\circ}$ inclination of the Cartwheel and
re-calculated for a zero inclination.

Both methods give very similar results. Angular differences in arcsec
between the peak positions are shown in Fig.~\ref{Fig11} as a function of
position angle, which is measured counterclockwise from a line extending
north from the outer ring center.  The {\it Ks}-band and {\it B}-band
peaks are located at smaller radii than the peak in $\rm H\alpha$ for almost
all position angles. The angular difference between the {\it Ks}-band and
$\rm H\alpha$ peaks is systematically bigger than between the {\it
B}-band and $\rm H\alpha$ peaks (Fig.~\ref{Fig11}b), which is
in full agreement with our model predictions.

Let us now consider the effects of finite thickness of the Cartwheel's
disk in more detail. Let us assume that the Cartwheel's disk has a radially
non-constant thickness, with the inner regions being thicker than the outer
regions. This assumption is justified, since the inner regions of the Cartwheel
are expected to be older and hence thicker than the outer regions. Internal
dust extinction is known to be considerable in the Cartwheel's outer ring
(Fosbury \& Hawarden \cite{FH}, VB). We assume
a limiting case when  part of the Cartwheel's outer ring
that is located below the galactic plane is totally unseen. Under these
assumptions, the angular differences between the peak positions should have
maximums along the minor semi-axis
that is drawn on the galaxy's near side.
The minima of angular differences
are then expected to be along the minor semi-axis drawn on the galaxy's
far side. We smoothed Fig.~\ref{Fig11} with a $30^\circ$ wide median filter 
in order to
minimize the noise. The results are shown by the solid lines in each panel
of Fig.~\ref{Fig11}. The angular differences between all three bands are
minimal along the southern minor semi-axis at position angles
180$^{\circ}$-220$^{\circ}$.
On the opposite side, along the northern minor semi-axis at position angles
20$^{\circ}$-40$^{\circ}$,
the angular differences in Fig.~\ref{Fig11} show an expected tendency to grow
and reach a maximum. This places the
southern part of the Cartwheel on the galaxy's far side.
Our conclusion is consistent with the analysis of Higdon (\cite{Higdon2}),
who deduced the galaxy's orientation indirectly from the form
of rotation curve and spokes. There is however an unexpected maximum in
the $H\alpha - K$ and $H\alpha -B$ angular peak differences at position angles
320$^{\circ}$-340$^{\circ}$, while the $B-K$ angular peak difference is minimal
there.  This may be a manifestation of an ${\rm H}\alpha$ warp in
the Cartwheel's outer ring.

As it has been noted above, differences between the peak positions measured
along the major axis are not influenced by the finite thickness of a ring
galaxy. Hence, they might be considered as an indicator of the Cartwheel's
outer ring propagation velocity.
The angular peak differences
along the Cartwheel's major axis determined from the smoothed curves
in Fig.~\ref{Fig11} are ${\rm H}\alpha-Ks$=1.97$^{\prime\prime}$,
${\rm H}\alpha-B$=1.29$^{\prime\prime}$, and $B-Ks$=0.67$^{\prime\prime}$.
They correspond to the linear peak differences
of ${\rm H}\alpha-Ks$=1.3~kpc, ${\rm H}\alpha-B$=0.85~kpc, and $B-Ks$=0.45~kpc
for the adopted distance of 140~Mpc.
These values are even bigger than those predicted by our
modeling for the Cartwheel's outer ring velocity of 90~km~s$^{-1}$,
implying that the outer ring is currently propagating at $v_{\rm ring}
\approx 110$~km~s$^{-1}$. We note that these
estimates cannot be considered as very accurate, though they clearly show
that the Cartwheel's outer ring is indeed expanding, and probably with
a considerable velocity as predicted by modeling of color gradients
in Sect.~\ref{gradients} and
numerical hydrodynamics modeling in Sect.~\ref{simulate}.  
We derive $v_{\rm ring} \approx 100$~km~s$^{-1}$ as an average of 
estimates obtained in Sect.~\ref{gradients}-\ref{surf_br}. Since the wave
velocity most probably decrease with time, we conclude that the Cartwheel
galaxy is no older than 250~Myr.

\section{Other possibilities for estimating the propagation velocity of ring
gas density waves}
\label{other}
\subsection{CaII triplet equivalent widths}

The CaII triplet (CaT) lines  originate in the atmospheres of cool stars
with the absorption being the strongest in red supergiants.  While
studying numerically the evolutionary behavior of the CaT equivalent widths
(EW(CaT)) in starbursts of different metallicities, Mayya (\cite{Mayya2})
found that besides the primary peak around 10~Myr the EW(CaT) show a well
defined secondary peak around 60~Myr in the solar metallicity starbursts.
The time separation $t_{\rm CaT}=50~Myr$ between the primary EW(CaT) peak
around 10~Myr and the secondary EW(CaT) peak around 60~Myr might be in
principle used to infer the propagation velocity of gas density waves in
ring galaxies.

We illustrate this possibility for the Cartwheel ring galaxy. We assume
that the Cartwheel's outer ring is currently propagating at $v_{\rm ring}$=90~km~s$^{-1}$.
For the time being, we assume that the Cartwheel's gaseous disk has the
solar metallicity. We also neglect the possible input to EW(CaT) from the
pre-collision stellar populations, which is indeed minimal at the
Cartwheel's outer parts (VB).  The filled
triangles in Fig.~\ref{Fig12} show the model radial profile of EW(CaT) after
the outer ring has propagated 24~kpc in the galactic disk.  Two well-defined peaks
of EW(CaT) at 19.1~kpc and 23.5~kpc are clearly evident.  For the outer
ring propagation velocity of
90~km~$s^{-1}$ it takes exactly 50~Myr to cross the distance between these
two peaks in Fig.~\ref{Fig12}. The angular difference between the peak
positions of 6.6" should be easily detected. Now given the linear difference
between the primary and secondary peak positions $R_{\rm CaT}$ is known from
observations of a ring galaxy, one can calculate the propagation velocity of
gas density wave by dividing $R_{\rm CaT}$ on the value of $t_{\rm
CaT}$. Or alternatively, if propagation velocity is known from independent
measurements or/and detailed modeling, one can calculate $R_{\rm CaT}$,
thus breaking new ground for the distance-independent estimates of a ring
galaxy linear diameter. Hence, EW(CaT) may provide the means for a
Hubble-constant-independent estimate of a distance to a ring galaxy.

\begin{figure}
  \resizebox{\hsize}{!}{\includegraphics{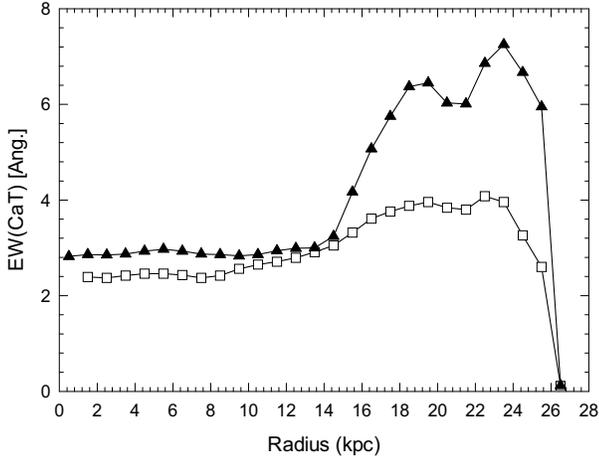}}
      \caption{The model radial profile of EW(CaT), as would be expected
      if the Cartwheel's disk had the solar metallicity (filled triangles)
      and $z_{\odot}/2.5$ metallicity (open squares).
              }
         \label{Fig12}
   \end{figure}

This method of evaluating the propagation velocity of gas density waves
is most effective in
the solar metallicity environments.  For sub-solar metallicities however the
primary and secondary peaks in EW(CaT) become much less pronounced as shown
by the open squares in Fig.~\ref{Fig12} for the $z_{\odot}/2.5$ metallicity.
In fact, the primary peak almost disappear for metallicities below
$z_{\odot}/5$. Furthermore, the absolute values of EW(CaT) generally
decrease for progressively lower metallicities (Mayya \cite{Mayya2}). These
difficulties make the observational detection of the primary and secondary
peaks problematic in the sub-solar metallicity environments. For the
Cartwheel galaxy, the solar metallicity is expected only in the central
parts. The outer parts are most probably of sub-solar $(z_{\odot}/3.5)$
metallicity and this method will not work.  Unfortunately, heavy element
abundance measurements of Bransford et al. (\cite{Bransford}) indicate that
individual star-forming knots in the rings of all ring galaxies in their
sample have sub-solar metallicities in the range from $z_{\odot}/2.5$ to
$z_{\odot}/5$.  We have to look for metal-rich ring galaxies in order to
apply this method for determining the propagation velocity of gas density wave.

\subsection{Other spectral lines}

Some other prominent absorption lines might be used for
estimating the propagation velocity of gas density waves in ring galaxies.
Synthesized spectra of a single-burst stellar population (PEGASE2 or Starburst99)
indicate that the equivalent widths of such optical absorption lines
as Mg~b, Na~D,  and some of strong Fe lines have a local maximum/minimum
between 10 and 250~Myr. A set of these lines can help to establish "time
labels" along the radius of a ring galaxy under an assumption that most stars
in the galaxy were born in a single star formation event.
Even two localizations of such extremes in the radial profiles of absorption
line equivalent widths
can provide us with the ages of stellar populations at different radii and
hence give us an insight into the value of propagation velocity of gas density wave
that triggered the star formation event. Use of a large aperture telescope 
in combination with a contemporary spectrograph makes it possible to 
obtain  high signal-to-noise spectra of ring galaxies in a few hours 
of integration time (Moiseev \cite{Moiseev}).

\subsection{Non-thermal radio emission}
Evolved stellar populations in the wake of an expanding ring density wave explode
as supernovae. Type~II supernova remnants are responsible for strong non-thermal
radio emission. Condon and Yin (\cite{CY}) argue that most Type~II
supernova remnants and their consequent non-thermal radio emission result from stars with
masses just over 8~$M_{\odot}$, while stars more massive than 20~$M_{\odot}$
account for most of the ionizing radiation and consequent $\rm H\alpha$ emission.
If so, we might expect to observe the difference  between the peak positions
of the radial  profiles of non-thermal radio emission
and $\rm H\alpha$ surface brightness in ring galaxies. Indeed, the ring galaxy VII~Zw~466
does show signs of such a difference when radio emission peaks on the inside
edge of the disk and does not coincide with the prominent HII region complexes.
The radial difference between the peaks in ${\rm H}\alpha$ and non-thermal radio
emission might be used to infer the propagation velocity of gas density
waves in ring galaxies in the same manner as in Sect.~\ref{surf_br}.

\section{Conclusions}
\label{sum}
Our numerical hydrodynamics modeling of ring galaxy formation demonstrates
that the maximum radial expansion velocity of gas ($v_{\rm gas}$) in the first ring
is always below the propagation velocity of the first ring itself ($v_{\rm
ring}$)
The inequality $v_{\rm gas} < v_{\rm ring}$ is less strict in ring
galaxies of comparable disk/halo masses than in halo-dominated ring
galaxies.  Numerical hydrodynamics modeling of the Cartwheel galaxy
indicates the outer ring is currently propagating at $v_{\rm ring} \approx
100$~km~s$^{-1}$, 
while the maximum radial expansion velocity of gas in
the outer ring is currently $v_{\rm gas}
\approx 65$~km~s$^{-1}$. The latter
value is in marginal agreement with the measurements of Higdon
(\cite{Higdon2}) based on HI kinematics, but substantially disagree with the
measurements of Amram et al. (\cite{Amram}) based on kinematics of ionized
gas. We conclude that the dynamical properties of neutral and ionized gas
may differ significantly in ring galaxies.

We demonstrate that the radial $B-V/V-K$ color gradients, together with the
$Q_{\rm BVK}$ extinction-free indices, can be used for estimating the value
of $v_{\rm ring}$ in ring galaxies.  Modeling of the
radial $B-V/V-K$ color gradients measured in the Cartwheel disk by Marcum et
al. (\cite{Marcum}) is sensitive to the assumed value of a Hubble constant.
Nevertheless, the color gradient modeling imposes the lower limit of 
$v_{\rm ring}=40$~km~s$^{-1}$ on the propagation velocity of the outer ring
for any value of a Hubble constant in the range 
$H_0=[50-100]$~km~s$^{-1}$~Mpc$^{-1}$. For the adopted in this work 
$H_0=65$~km~s$^{-1}$~Mpc$^{-1}$, the best agreement between
the model and observed radial color gradients is  achieved for 
velocities of 90~km~s$^{-1}$ and even higher. 

Our modeling of the Cartwheel galaxy demonstrates that the angular differences
between the peaks in the $\rm H\alpha$, {\it K-}, and {\it B}-band
radial surface brightness profiles are
${\rm H\alpha}-K$=1.65$^{\prime\prime}$ and ${\rm H\alpha}-B$=1.05$^{\prime\prime}$,
which should
be detected with the achieved resolution. On the other hand, we find that
the azimuthally averaged radial $\rm H\alpha$ surface brightness profile of
the Cartwheel's outer ring does not peak exterior to those in {\it Ks-} and
{\it B}-bands, contrary to the predictions of our model for $v_{\rm ring} = 90$~km~s$^{-1}$.
We conclude that a 41$^\circ$ inclination, along with a finite thickness and
warping of the Cartwheel's disk, act to smear out the angular differences
between the peaks in the $ \rm H\alpha$, {\it Ks}, and {\it B} radial profiles.
Indeed, we construct a series of the radial surface brightness profiles azimuthally
averaged over a 1$^\circ$ sector. These pie-chart-like radial profiles do show
an expected tendency and the ${\rm H}\alpha$ radial profile
peaks exterior to those in {\it Ks-} and {\it B}-bands at the position of the
Cartwheel's outer ring. The ${\rm H\alpha}-Ks$ and ${\rm H\alpha}-B$ peak
differences are the biggest along the Cartwheel's major
axis, where effects of inclination and finite thickness are minimized,
and are in close agreement with our model predictions.
The peak differences obtained along the Cartwheel's major axis 
imply $v_{\rm ring} \approx ~$~110~km~s$^{-1}$.
Azimuthal variations in ${\rm H}\alpha-Ks$, ${\rm H}\alpha-B$, and $B-Ks$
peak differences support our assumptions about a thick and non-flat nature
of the Cartwheel's outer ring.
We conclude that the Cartwheel's 
outer ring is currently propagating at $v_{\rm ring} \approx  100$~km~s$^{-1}$,   
which is in agreement with our numerical hydrodynamics results.

We show that the equivalent widths of some spectral lines, such as the
CaII triplet, might be used for
determining the propagation velocity of  ring density waves in {\it solar metallicity
environments}. Solar metallicity rings have not been detected so far (Bransford
et al. \cite{Bransford}), but they may be expected in young ring galaxies
where the rings have not yet expanded out to the metal-unpolluted regions.   
Non-thermal radio emission is also a promising tool for
determining the wave velocity in ring galaxies with high rates of star
formation like the Cartwheel.

\begin{acknowledgements}
We are thankful to the referee, Dr. A.P. Marston, for his suggestions
 and critical comments, which helped us to improve the paper. 
The authors are grateful to Dr. Y.D. Mayya for providing his population
synthesis code. We thank Dr. S.Popov for help with data collection.

The paper is based on observations made with ESO Telescopes at the Paranal
Observatory under program ID 66.B-0666(B).

Some of the data presented in this paper were obtained from the
Multimission Archive at the Space Telescope Science Institute (MAST). STScI
is operated by the Association of Universities for Research in Astronomy,
Inc., under NASA contract NAS5-26555. Support for MAST for non-HST data is
provided by the NASA Office of Space Science via grant NAG5-7584 and by
other grants and contracts.

The paper is based on archival data taken from the Canadian Astronomy Data Center,
which is operated by the Dominion Astrophysical Observatory for the National
Research Council of Canada's Herzberg Institute of Astrophysics.

\end{acknowledgements}

\end{document}